\documentstyle[eqsecnum,preprint,aps,prd]{revtex}
\preprint{WISC-MILW-94-TH-16 $\qquad \qquad$ FERMILAB-Pub-94/351-A}
\tighten 
\begin{document} \draft \title{CBR Temperature
Fluctuations Induced by Gravitational Waves in a Spatially-Closed
Inflationary Universe} \author{Bruce Allen} \address{ Department of
Physics, University of Wisconsin -- Milwaukee\\ P.O. Box 413,
Milwaukee, Wisconsin 53201, U.S.A.\\ email:
ballen@dirac.phys.uwm.edu} \author{Robert Caldwell}
\address{NASA/Fermilab Astrophysics Center\\ Fermi National
Accelerator Laboratory\\ P.O. Box 500, Batavia, Illinois 60510\\
email: caldwell@astro1.fnal.gov} \author{Scott Koranda} \address{
Department of Physics, University of Wisconsin -- Milwaukee\\ P.O.
Box 413, Milwaukee, Wisconsin 53201, U.S.A.\\ email:
skoranda@dirac.phys.uwm.edu} \date{\today} \maketitle
\begin{abstract} Primordial gravitational waves are created during
the de Sitter phase of an exponentially-expanding (inflationary)
universe, due to quantum zero-point vacuum fluctuations.  These waves
produce fluctuations in the temperature of the Cosmic Background
Radiation (CBR).   We calculate the multipole moments of the
correlation function for these temperature fluctuations in a
spatially-closed Friedman-Robertson-Walker (FRW) cosmological model.
The results are compared to the corresponding multipoles in the
spatially-flat case.  The differences are small unless the density
parameter today, $\Omega_0$, is greater than 2.  \end{abstract}
\section{INTRODUCTION} \label{section1} Inflationary models of the
early universe contain a well-studied mechanism which creates
primordial fluctuations.  The fluctuations originate as
quantum-mechanical zero-point fluctuations during the
exponentially-expanding de Sitter phase.  By a process which may be
variously described as particle (graviton) production, non-adiabatic
amplification, or super-radiant scattering, these fluctuations become
large in the present epoch.  As the universe expands, these
perturbations are redshifted to longer wavelengths and amplified;
during the present epoch these perturbations typically persist over a
range of wavelengths $\lambda$ from $10^{-27} {\rm \> cm} < \lambda <
10^{2} {\rm \> cm}$.  For a review of perturbations in inflationary
models, see Kolb and Turner \cite{KolbTurner}.

The perturbations of the gravitational field may be decomposed into
scalar, vector and tensor components.  The tensor perturbations
considered in this paper may be thought of as gravitational waves in
a classical description, or as spin-two gravitons in the quantum
mechanical description used in the present work.  The modes of
interest have present-day frequencies in the range from $10^{-17}$ Hz
to $10^{-12}$ Hz and have extremely large occupation numbers.  Hence
they may also be thought of as classical gravitational waves - the
two descriptions coincide.  The gravitons are created during the de
Sitter phase of rapid expansion by the mechanism originally proposed
by Parker;  the same mechanism creates particles near a black hole or
in any other region where the spacetime curvature is large and
particle creation is not forbidden by global symmetries or
conservation laws.  A simple calculation showing how a
potentially-observable spectrum of gravitons is created in inflation
is given by Allen \cite{Allen}.

The tensor perturbations of the gravitational field produce
temperature fluctuations in the CBR, via the Sachs-Wolfe effect.  The
expected values of the resulting temperature fluctuations are
described by the angular correlation function \begin{equation}
C(\gamma) \equiv\bigg\langle 0\bigg| {{\delta T} \over
T}(\Omega){{\delta T}\over T}(\Omega')\bigg|0\bigg\rangle=
\sum_{l=1}^\infty {(2l+1)\over 4\pi}
 {\langle a_l^2 \rangle } P_l(\cos\gamma).  \label{cofgamma}
\end{equation} Here $\delta T/T(\Omega)$ is the fractional
temperature fluctuation in the CBR at point $\Omega$ on the
observer's celestial sphere, $\gamma$ is the angle between $\Omega$
and $\Omega'$, and the quantum expectation value is evaluated in the
initial state of the universe.  The multipole moments $ {\langle
a_l^2 \rangle } $ are generally used to describe $C(\gamma)$.

In a recent paper \cite{AllenKoranda} the expected multipole moments
$ {\langle a_l^2 \rangle } $ due to tensor perturbations are
calculated in a spatially flat $k=0$ FRW inflationary model.  That
paper contains a detailed review of previous work on this problem, a
comprehensive description of the physical motivation, and a detailed
and self-contained ``first-principles'' calculation.  The present
work repeats that calculation in the spatially-closed ($k=+1$) case.
The only previous work on tensor perturbations in the
spatially-closed case is that of Abbott and Schaefer
\cite{AbbottSchaefer}.  Note that the angle brackets around $a_l^2 $
serve as a reminder that we are calculating the {\it expected} or
{\it expectation values} of these multipole moments, not necessarily
the values that they might have in any given realization of the
universe.

The calculation in this paper follows the previous work by Allen and
Koranda \cite{AllenKoranda} very closely.  In the present work, we
will assume that the reader is familiar with that earlier paper, and
present only the bare minimum of detail required to generalize the
work to the $k=+1$ case.  In Section \ref{section2} we present the
$k=+1$ cosmological model and Sachs-Wolfe effect.  Section
\ref{section3} gives the form of the metric perturbation operator for
linearized gravitational fluctuations.  Section \ref{section4}
combines these results to obtain an analytic form for the multipole
moments ${\langle a_l^2 \rangle }$.  Section \ref{NumEval} details
the method by which these multipole moments were evaluated
numerically, and Section \ref{section6} outlines the results and
conclusions of that numerical study.

Throughout this paper, we use units where the speed of light $c=1$.
However for clarity we have retained Newton's gravitational constant
$G$ and Planck's constant $\hbar$ explicitly.  We choose function
branches so that $\sqrt{x} \ge 0$ and $\arcsin(x) \in
[-\pi/2,\pi/2]$.


\section{The Background Space-Time and the Sachs-Wolfe Effect}
\label{section2} The spacetime considered here has the topology $R
\times S^3$ of the static Einstein cylinder, and is covered by
coordinates $x^0=t,x^1=\chi, x^2=\theta, x^3=\phi$ with the ranges
$\chi,\theta \in [0,\pi]$, and $\phi \in [0, 2\pi)$.  The time
coordinate $t$ ranges over a connected open subset of the real line,
which we will specify below.  The spatial coordinates cover a
three-sphere of radius $a(t)$; we refer to this function as the
cosmological scale factor.  The metric of the spacetime is given by
\begin{equation} $$ds^2 = a^2(t) \biggl( -dt^2 + d\chi^2 + \sin^2
\chi (d\theta^2 + \sin^2 \theta d\phi^2) + h_{ij}(t,\chi,\theta,\phi)
dx^i dx^j \biggr).  \end{equation} The metric perturbation $h_{ij}$
is assumed to be small; in its absence the spacetime metric is that
of a homogeneous and isotropic $k=+1$ FRW model.  With our choice of
gauge for the tensor metric perturbations, the indices $i,j = 1,2,3$
run only over the spatial coordinates.

In order to completely specify the cosmological model, we need to
define the cosmological scale-factor $a(t)$.  The cosmological model
is completely defined by the free parameters given in Table
\ref{FreeParameterTable}.  Note that we have assumed that the
universe is currently expanding, since we require $H_0$ to be
positive.  The density parameter \begin{equation} \Omega_0 =  {8 \pi
G \rho_0 \over 3 H_0^2} \end{equation} is the ratio of the
present-day energy-density $\rho_0$ to the critical energy density
required to produce a spatially-flat $k=0$ universe.

\subsection{The Matter-Dominated (Dust) Phase} In our cosmological
model, the universe is assumed to pass through three ``phases'',
appropriate to a simple inflationary model.  We let $t=0$ denote the
present time.  The most recent phase was a matter-dominated period of
expansion, described by the scale factor \begin{equation} a(t) = A
\sin^2(t/2 + B) \qquad {\rm for} \quad t_{\rm eq} < t <0.
\end{equation} Here the constants $A,B,t_{\rm eq}$ are defined by
\begin{eqnarray} A & = & {1 \over H_0} \Omega_0 (\Omega_0-1)^{-3/2},
\nonumber \\ B & = & \arcsin \sqrt{\Omega_0-1 \over \Omega_0}, {\rm
\quad and} \label{Teq}
 \\ t_{\rm eq} & = & 2  \arcsin \sqrt{\Omega_0-1 \over \Omega_0
(1+Z_{\rm eq}) }
 -2 \arcsin \sqrt{\Omega_0-1 \over \Omega_0}. \nonumber
\end{eqnarray} During this matter-dominated phase, the stress-energy
tensor is that of a perfect fluid, with zero pressure and an energy
density proportional to $a^{-3}(t)$.  We assume (as indicated in
Table \ref{FreeParameterTable}) that the surface of last scattering
is located within the matter-dominated phase.   Thus, the time of
last scattering, \begin{equation} t_{\rm ls}  =  2  \arcsin
\sqrt{\Omega_0-1 \over \Omega_0 (1+Z_{\rm ls}) }
 -2 \arcsin \sqrt{\Omega_0-1 \over \Omega_0}, \end{equation} is given
by a formula identical in form to (\ref{Teq}) for $t_{\rm eq}$, and
satisfies $t_{\rm eq}<t_{\rm ls} < 0$.

\subsection{The Radiation-Dominated Phase} Preceding the
matter-dominated phase of expansion is a radiation-dominated phase of
expansion.  During this phase the scale factor is \begin{equation}
a(t) = C \sin(t + D) \qquad {\rm for} \quad   t_{\rm end} < t <
t_{\rm eq} .  \end{equation} Here the constants $C,D,t_{\rm end}$ are
defined by \begin{eqnarray} C & = & {1 \over H_0} \Omega_0^{1/2}
(\Omega_0-1)^{-1}  (1+Z_{\rm eq})^{-1/2}, \nonumber \\ D & = & 2
\arcsin \sqrt{\Omega_0-1 \over \Omega_0} - \arcsin \sqrt{\Omega_0-1
\over \Omega_0 (1+ Z_{\rm eq})}, {\rm \quad and}\\ t_{\rm end} & = &
- 2 \arcsin \sqrt{\Omega_0-1 \over \Omega_0} +\arcsin
\sqrt{\Omega_0-1 \over \Omega_0 (1+Z_{\rm eq}) } + \arcsin
\sqrt{(\Omega_0-1) (1+Z_{\rm eq}) \over \Omega_0 (1+Z_{\rm end})^2
}.  \nonumber \end{eqnarray} During this radiation-dominated phase of
expansion the energy density is proportional to $a^{-4}(t)$ and the
pressure is equal to $1/3$ of the energy-density.  This phase is
preceded by a de Sitter phase.

\subsection{The Initial de Sitter (Inflationary) Phase} In our
coordinate system, the de Sitter (exponentially expanding,
inflationary) phase has scale factor \begin{equation} a(t) = {E
\over  \sin(t + F)} \qquad {\rm for} \quad  t < t_{\rm end} .
\end{equation} Here the constants $E$ and $F$ are defined by
\begin{eqnarray} E & = & {-1 \over H_0} \Omega_0^{-1/2} (1+Z_{\rm
eq})^{1/2} (1+Z_{\rm end})^{-2}, {\rm \quad and}\\ F & = & 2 \arcsin
\sqrt{\Omega_0-1 \over \Omega_0} -\arcsin \sqrt{\Omega_0-1 \over
\Omega_0 (1+ Z_{\rm eq})} -2  \arcsin \sqrt{(\Omega_0-1) (1+Z_{\rm
eq}) \over
 \Omega_0 (1+Z_{\rm end})^2 }.  \nonumber \end{eqnarray} Note that
the constant $E<0$ because $\sin(t+F)<0$ during the de Sitter phase.
During the de Sitter phase, the energy density is a constant
\begin{equation} \rho_{\rm de Sitter} = {3 \over 8 \pi} E^{-2} =
\rho_0 {(1+Z_{\rm end})^4 \over 1+Z_{\rm eq}} = {3 H_0^2 \Omega_0
\over 8 \pi G} {(1+Z_{\rm end})^4 \over 1+Z_{\rm eq}} \end{equation}
and the (negative) pressure is $-\rho_{\rm de Sitter}$.

\subsection{Properties of the Cosmological Model} It may be easily
verified that the scale factor and its derivative w.r.t. time $t$ are
both continuous, however the second derivative is discontinuous.
This is because in our simple inflationary model, the energy-density
is a continuous function but the pressure changes discontinuously at
the beginning and end of the radiation-dominated epoch.

The de Sitter phase ``begins'' at early times when the time
coordinate $t $ approaches the value $ -\pi - F$.  At this early time
the cosmological scale factor is very large (approaching infinity as
$t \to -\pi - F$).   As the time coordinate increases, the scale
factor decreases, eventually reaching a minimum value when $t =
t_{\rm min}=-\pi/2 - F$.  After this time, the scale factor begins to
increases again (exponentially in physical time).  One might find it
reasonable to demand that the universe be expanding at time $t_{\rm
end}$ when the inflationary phase ends.  This is the case if and only
if $t_{\rm min} < t_{\rm end}$, which implies that the free
parameters given in Table \ref{FreeParameterTable} must satisfy the
inequality \begin{equation} \sqrt{ (\Omega_0-1) (1+Z_{\rm eq}) \over
\Omega_0 } < 1 + Z_{\rm end}.  \label{ExpandingAtEnd} \end{equation}
It is also easy to determine the ``amount'' of inflation that takes
place.  The amount that the universe has expanded between time
$t_{\rm min}$, when the spatial sections have their smallest extent,
and time $t_{\rm end}$, when the inflationary phase terminates and
the radiation-dominated phase begins, is \begin{equation} {a(t_{\rm
end}) \over a(t_{\rm min})} = (1+Z_{\rm end}) \sqrt{\Omega_0 \over
(\Omega_0-1)(1+Z_{\rm eq})}.  \label{AmountOfInflation}
\end{equation} Comparison with (\ref{ExpandingAtEnd}) shows the
obvious - if the universe is expanding at the end of the de Sitter
phase, then the amount of inflationary expansion
(\ref{AmountOfInflation}) is greater than unity.  In typical
inflationary models, the free parameters have values of order $H_0$
between $50$ and $100$ Km/s-Mpc, $ \Omega_0 < 2$, $100< Z_{\rm ls} <
1500$, $2 \times 10^3 < Z_{\rm eq} < 2 \times 10^4$, and $10^{26} <
Z_{\rm end}$.

There is a sense in which the spatially-closed inflationary models
are not ``natural."  One of the principal motivations which led to
the development of the inflationary paradigm was the desire to solve
the so-called ``horizon problem."  As we will now show, this problem
is only solved (for reasonable choices of the cosmological
parameters) if $\Omega_0<2$.  Thus, while it is technically
consistent to use the results obtained in this paper for any value of
$\Omega_0>1$, one must bear in mind that the cosmological model, for
large values of $\Omega_0$, runs counter to the spirit of inflation.

The horizon problem may be stated in terms of a set of points {\cal
C}, which is the intersection of the past horizon of an observer
today with the surface of last scattering. The horizon problem is
``solved" if {\cal C} lies within the causal domain of influence of
either (1) a point on the initial singularity, in a big bang model,
or (2) a point on the surface at $t=t_{\rm min}$ where inflation
``begins", in a model with no initial singularity.  Thus, in our
model, which is of type (2), the horizon problem is solved if and
only if \begin{eqnarray} &|t_0 - t_{\rm ls}| < |t_{\rm ls} - t_{\rm
min}|& \nonumber \\ & \iff  & \\ & 2 \arcsin \sqrt{\Omega_0-1 \over
\Omega_0} -4 \arcsin \sqrt{\Omega_0-1 \over \Omega_0 (1+ Z_{\rm ls})}
+ \arcsin \sqrt{\Omega_0-1 \over \Omega_0 (1+ Z_{\rm eq})} +2
\arcsin \sqrt{(\Omega_0-1) (1+Z_{\rm eq}) \over \Omega_0 (1+Z_{\rm
end})^2 } < {\pi \over 2}. & \nonumber \end{eqnarray} For reasonable
cosmological models, the terms containing $Z_{\rm ls}$, $Z_{\rm eq}$,
and $Z_{\rm end}$ may be neglected.  The horizon problem is then
solved if and only if \begin{equation} \arcsin \sqrt{\Omega_0-1 \over
\Omega_0} < {\pi \over 4} \iff \Omega_0 < 2.  \end{equation} While we
present results for any value of $\Omega_0$, the cosmological model
itself should be viewed with some suspicion if $\Omega_0$ is much
larger than unity.

The final result of this paper are values of the dimensionless
quantities \begin{equation} M_l \equiv {\rho_{\rm Planck} \over
\rho_{\rm de Sitter}} { l(l+1) \over 6}
   {\langle a_l^2 \rangle } .  \end{equation} Here $\rho_{\rm
Planck}$ is the Planck energy-density $\rho_{\rm Planck} = {1 \over
\hbar G^2} \approx 5 \times 10^{93} \>\rm gm/cm^3$.  It will turn out
that $M_l$ is independent of $H_0$, and depends only upon the
dimensionless quantities $\Omega_0$, $Z_{\rm ls}$, $Z_{\rm eq}$, and
$Z_{\rm end}$.  This is because $\rho_{\rm de Sitter}$, and $
{\langle a_l^2 \rangle }  $ are both proportional to $H_0^2$.  In
addition, if $Z_{\rm end}$ is sufficiently large then the $M_l$ are
also independent of its value.

\subsection{The Sachs-Wolfe Effect} If the metric perturbation
$h_{ij}$ vanishes and the temperature of the CBR on the surface of
last scattering is constant, an observer today would see exactly the
same temperature at each point on the celestial sphere, and
$C(\gamma)$ would vanish.  However the metric perturbations will in
general break the rotational symmetry and perturb the energy of the
photons.  This results in a temperature fluctuation which varies from
point to point on the celestial sphere; the fluctuation may be
calculated in the same way as for a spatially-flat Universe, given in
\cite{SachsWolfe}.

We assume that the observer is located at $t=0$, and at ``radial''
coordinate $\chi=0$.  (Because the coordinate system is singular at
$\chi=0$ every value of $\theta,\phi$ corresponds to the same
space-time point at $\chi=0$, so their values are irrelevant when
$\chi=0$.)  If the observer looks out at a point $\Omega$ on the
celestial sphere, she observes photons that arrive, in the
unperturbed metric, along the null geodesic path \begin{equation}
t(\lambda) = \lambda \quad \chi(\lambda) = |\lambda| = -\lambda \quad
\theta(\lambda) = \theta_{\Omega} \quad \phi(\lambda) =
\phi_{\Omega}.  \end{equation} In these equations, $ \theta_{\Omega}$
and $\phi_{\Omega}$ are the angular coordinates of the point $\Omega$
on the celestial two-sphere.  We have chosen the (non-affine)
parameter $\lambda$ along the null geodesic path to run through the
range $t_{\rm ls} \le \lambda \le 0$ between the time of last
scattering and the observation today.

In the presence of the metric perturbation $h_{ij}$ the fractional
temperature fluctuation observed at point $\Omega$ on the celestial
sphere is \begin{equation} {\delta T \over T}(\Omega) = - {1 \over 2}
\int_{t_{\rm ls}}^0 d\lambda \> \biggl({\partial h_{\chi \chi} \over
\partial t}
\biggr)({t(\lambda),\chi(\lambda),\theta(\lambda),\phi(\lambda)}).
\label{SachsWolfeFormula} \end{equation} As indicated in this
formula, the partial derivative w.r.t. the time coordinate $t$ is
taken before setting the coordinates equal to the values which they
take along the unperturbed null geodesic path.

\section{The Metric Perturbation Operator} \label{section3} The
classical metric perturbation $h_{ij}$ may be replaced with a quantum
field operator.  The justification for this is given in our detailed
paper on the spatially-flat case \cite{AllenKoranda} and will not be
repeated here.  The basic idea is that the inflationary epoch
redshifts away all the perturbations, with the exception of the
zero-point quantum fluctuations.  Hence we calculate the expectation
value of ${{\delta T} \over T}(\Omega){{\delta T}\over T}(\Omega')$
in the vacuum state $\bigl| 0 \rangle$ appropriate to the initial de
Sitter state.

\subsection{Mode function expansion of Metric Perturbation Operator}
The quantum field operator (which we denote with the same symbol
$h_{ij}$ as the corresponding classical perturbation) may be expanded
in terms of a complete set of mode functions.  As was originally
shown by Ford and Parker \cite{FordParker}, in an FRW cosmological
model, the time-dependence of these mode functions is the same as
that of a massless minimally-coupled scalar field.  The field
operator is \begin{eqnarray} \label{FieldOp}
h_{ij}(t,\chi,\theta,\phi) =& \\ \sum_{L=2}^\infty \sum_{l=2}^L
\sum_{m=-l}^{l} \biggl[ & \psi_{Llm}(t)
T^{(s;Llm)}_{ij}(\chi,\theta,\phi) c_{Llm} + \psi^*_{Llm}(t)
T^{*(s;Llm)}_{ij}(\chi,\theta,\phi) c_{Llm}^\dagger \nonumber \\ + \>
& \psi_{Llm}(t) T^{(v;Llm)}_{ij}(\chi,\theta,\phi) d_{Llm} +
\psi^*_{Llm}(t) T^{*(v;Llm)}_{ij}(\chi,\theta,\phi) d_{Llm}^\dagger
\biggr]. \nonumber \end{eqnarray} In this expression, the sum is over
a complete set of rank-two symmetric transverse traceless tensors
$T^{(Llm)}_{ij}$.  These tensor modes are defined on a unit-radius
sphere $S^3$ and are given explicitly by Higuchi \cite{Higuchi2005}.
(Note however a typo \cite{HiguchiTypo} in one of the formulae which
does not affect the results which we need.) Henceforth we will denote
the triple sum that appears in (\ref{FieldOp}) by $\sum_{Llm}$
without explicitly indicating the ranges of summation.  (The
temperature $T$ can always be distinguished from the tensor modes
$T^{(Llm)}_{ij}$, since the latter is always written with indices.)

The graviton has two possible polarization states, labeled $s$ and
$v$ in this expansion, each of which has its own set of tensor
modes.  The modes are labeled by the three integers $L$, $l$, and
$m$.  (Note that Ford and Parker's \cite{FordParker} index $n=L+1$ in
our notation.) Associated with the $s$-polarization modes are
creation and annihilation operators $c_{Llm}$ and $c_{Llm}^\dagger$,
and associated with the $v$-polarization modes are creation and
annihilation operators $d_{Llm}$ and $d_{Llm}^\dagger$.  The only
non-vanishing commutation relation among this infinite set of
operators is the relation \begin{equation}
[c_{Llm},c_{L'l'm'}^\dagger] = [d_{Llm},d_{L'l'm'}^\dagger] =
\delta_{LL'} \delta_{ll'} \delta_{mm'} \label{Commutator}
\end{equation} where $\delta$ denotes the Kronecker delta function.

\subsection{The Transverse-Traceless-Symmetric Tensor Harmonics} The
tensor modes defined by Higuchi \cite{Higuchi2005} obey the
normalization condition \begin{equation} \int_0^\pi d\chi \sin^2 \chi
\int_0^\pi d\theta  \sin \theta \int_0^{2 \pi} d\phi T^{(p;Llm)}_{ij}
T^{(p';L'l'm')}_{rs} P^{ir} P^{js} = \delta_{LL'} \delta_{ll'}
\delta_{mm'} \delta_{pp'}.  \label{ModeNormalization} \end{equation}
Here, the polarization indices $p$ and $p'$ take on either of the
values ``s'' or ``v''.  The integral is over the unit-radius
three-sphere, and $P^{ir}$ is the inverse of the metric on the
unit-radius three-sphere:  \begin{equation} P_{ij} dx^i dx^j =
d\chi^2 + \sin^2 \chi (d\theta^2 + \sin^2 \theta d\phi^2).
\label{UnitSphereMetric} \end{equation} Note that the measure that
appears in the normalization integral (\ref{ModeNormalization}) is
the usual volume element defined by $\sqrt{\det P_{ij} }$.  The
Sachs-Wolfe effect (\ref{SachsWolfeFormula}) is produced only by the
$\chi \chi$ component of $h_{ij}$.  Because $T^{(v;Llm)}_{\chi\chi}
\equiv 0$, only the ``s'' polarization state contributes to the
temperature fluctuation.  The only component needed is thus
\begin{equation} T_{\chi \chi}^{(s;Llm)} (\chi,\theta,\phi) \equiv
{R_L^l}(\chi) Y_{lm}(\theta,\phi).  \end{equation} The
$Y_{lm}(\theta,\phi)$ are standard scalar spherical harmonic
functions on the two-sphere \cite{AbramowitzStegun}.  The ``radial''
dependence is given by \begin{equation} {R_L^l}(\chi) \equiv \sqrt{
(l-1)l(l+1)(l+2)(L+1)(L+l+1)! \over 2 L (L+1)^2 (L+2) (L-l)!} (\sin
\chi)^{-5/2} \> {\rm P}^{-(l+1/2)}_{L+1/2}(\cos \chi),
\label{RadialFun} \end{equation} where the functions ${\rm
P}^{-(l+1/2)}_{L+1/2}(z)$ are associated Legendre functions
\cite{AbramowitzStegun}.  In Section \ref{NumEval} we explain how
these functions may be easily evaluated.

\subsection{Normalization Condition for Wavefunctions} The quantum
field operator $h_{ij}$ obeys canonical commutation relations which
can be derived from the quadratic part of the gravitational action.
We have already specified the normalization of the
creation/annihilation operators (\ref{Commutator}) and of the spatial
part of the mode functions (\ref{ModeNormalization}).  The
commutation relations for $h_{ij}$ then determines the normalization
of the time part $\psi_{Llm}(t)$ of the graviton wavefunctions.  The
details of this procedure are given in \cite{FordParker} and yield a
normalization condition \begin{equation} \psi_{Llm}(t) {d \over dt}
\psi^*_{Llm}(t) - \psi^*_{Llm}(t) {d \over dt} \psi_{Llm}(t) = {32 i
\pi \hbar G} a^{-2}(t).  \end{equation} (Note that the normalization
condition given in equation (3.3) of Ford and Parker
\cite{FordParker} contains a minor typo \cite{FordParkerTypo}.)

\subsection{Choice Of An Initial (Vacuum) State} If one defines a
Fock vacuum state by the property that it is annihilated by all of
the operators $c_{Llm}$ and $d_{Llm}$, then the choice of vacuum
state is really determined by the choice of the mode functions
$\psi_{Llm}(t)$.  For the reasons given in \cite{AllenKoranda} we
choose these mode functions to be those which correspond to the
unique de Sitter invariant vacuum state $\bigl| 0 \rangle$ during the
initial inflationary stage whose two-point function has Hadamard
form.

\subsection{Wavefunction During The De Sitter Phase} As shown in
equation (2.18) of Ford and Parker \cite{FordParker}  the mode
functions obey the minimally-coupled massless scalar wave equation
\begin{equation} \biggl[ {d^2 \over dt^2} + {2 \over a} {d a \over
dt} {d \over dt} + L(L+2) \biggr] \psi_{Llm}(t) =0.
\label{ScalarWaveEqn} \end{equation} There is a slight subtlety: it
is impossible to define a de Sitter invariant Fock vacuum state for
the minimally-coupled massless scalar field \cite{AllenVacStates}.
However it was shown by Allen and Folacci \cite{AllenFolacci} that
the difficulty only arises for the $L=0$ mode.  In the case of the
gravitational field operator, the $L=0$ and the $L=1$ modes are both
absent; they correspond to ``monopole'' and ``dipole'' dynamical
degrees of freedom which are not present in the spin-two case.  Hence
in the case of the gravitational field, it is possible to define the
desired de Sitter invariant vacuum state.  The corresponding
normalized wavefunction during the de Sitter phase is
\cite{Allen,Higuchi2005,AllenVacStates,AllenFolacci,TagirovChernikov,Higuchi}
\begin{equation} \psi_{Llm}(t) = \psi_L(t)={1 \over a} \sqrt{16 \pi
\hbar G \over L(L+1)(L+2)} \biggl( i(L+1) - {1 \over a} {da \over dt}
\biggr) e^{-i (L+1) t} \quad {\rm for} \quad t<t_{\rm end}.
\end{equation} We note in passing that this time-dependent part of
the wavefunction depends only upon $L$ and not upon $l$ and $m$.
(This guarantees that the vacuum state will be invariant under all
rotations of the three-sphere $t=\rm constant$, which is the subgroup
$SO(4)$ of the de Sitter group $SO(1,4)_0$.  However the invariance
of the state under the de Sitter group $SO(1,4)_0$ is {\it not}
obvious from inspection.  Here the subscript on $SO(1,4)$ denotes the
part of the group connected to the identity.)  For this reason, from
this point on we drop the indices $l,m$ from the time-dependent part
of the wavefunction, denoting $\psi_{Llm}$ by $\psi_L$.  Because the
wave equation (\ref{ScalarWaveEqn}) is a second-order ODE, the
solution $\psi_L(t)$ during the de Sitter phase completely determines
the solution at all later times.  The solution at later times is
conveniently written in terms of Bogoliubov coefficients.

\subsection{Wavefunction During The Radiation-Dominated Phase} The
epoch that follows the de Sitter epoch is the radiation-dominated
phase.  One may write the solution to the wave equation during this
phase as \begin{equation} \psi_L(t) = \alpha^{rad}_L \psi^{rad}_L(t)
+ \beta^{rad}_L \psi^{*rad}_L(t) \quad {\rm for} \quad t_{\rm end} <
t < t_{\rm eq}.  \label{RadModeFun} \end{equation} Here, the positive
frequency mode during the radiation epoch is defined by
\begin{equation} \psi^{rad}_L(t) = {1 \over a} \sqrt{16 \pi \hbar G
\over (L+1)} e^{-i (L+1) t} \quad {\rm for} \quad t_{\rm end} < t <
t_{\rm eq}.  \end{equation} The Bogoliubov coefficients are
determined by a condition which follows from the wave equation
(\ref{ScalarWaveEqn}):  both $\psi_L(t)$ and its time derivative must
be continuous at all times, and in particular at $t=t_{\rm end}$.
One obtains \begin{eqnarray} \label{RadiationBog} \alpha^{rad}_L & =
&(L(L+2))^{-1/2} \biggl( i(L+1) + \sqrt{Q-1} - {iQ \over 2 (L+1)}
\biggr)\\ \beta^{rad}_L & = & {i \over 2} (L+1)^{-1} (L(L+2))^{-1/2}
Q e^{-2 i (L+1) t_{\rm end}}. \nonumber \end{eqnarray} Here $Q$ is
the constant defined by \begin{equation} Q= {\Omega_0 (1+Z_{\rm
end})^2 \over (\Omega_0-1)(1+Z_{\rm eq})}.  \label{QDef}
\end{equation} We stress once again that {\it the solution
$\psi_L(t)$ during the de Sitter phase completely determines the
solution at all later times}.  In other words {\it the choice of a
``positive-frequency'' mode function during the radiation phase is
unimportant.}  Had we picked a different solution to the wave
equation (\ref{ScalarWaveEqn}) to call ``positive frequency'' then
$\alpha^{rad}_L$ and $\beta^{rad}_L$ would have changed in such a way
as to keep the mode function $\psi_L(t)$ given in (\ref{RadModeFun})
unchanged.  In similar fashion, the solution of the wave equation
during the radiation phase completely determines its solution during
the matter-dominated phase.

\subsection{Wavefunction During The Matter-Dominated Phase} The
wavefunction during the matter-dominated (dust) phase may again be
expressed as a linear combination of the natural positive-frequency
solution and its complex conjugate:  \begin{equation} \psi_L(t) =
\alpha_L \psi^{mat}_L(t) + \beta_L \psi^{*mat}_L \quad {\rm for}
\quad t_{\rm eq} < t < 0.  \label{FinalModeFunction} \end{equation}
The positive frequency mode functions during the matter epoch are
\begin{equation} \psi^{mat}_L(t) = {1 \over a} \sqrt{16 \pi \hbar G
\over (L+1)(2L+1)(2L+3) } \biggl( 2i(L+1) +{1 \over a} {da \over dt}
\biggr) e^{-i (L+1) t} \quad {\rm for} \quad t_{\rm eq} < t < 0.
\label{PosFreqMat} \end{equation} The Bogoliubov coefficients
$\alpha_L$ and $\beta_L$ are determined (as in the spatially flat
case \cite{AllenKoranda}) by combining the Bogoliubov coefficients
for the two different phases.  \begin{equation} \pmatrix{\alpha_L
&\beta_L\cr \beta_L^* &\alpha_L^*}= \pmatrix{\alpha_L &\beta_L\cr
\beta_L^* &\alpha_L^*}^{\rm rad} \pmatrix{\alpha_L &\beta_L\cr
\beta_L^* &\alpha_L^*}^{\rm mat}.  \label{alphabetadefinition}
\end{equation} As previously, the Bogoliubov coefficients
$\alpha^{mat}_L$ and $\beta^{mat}_L$ are determined by matching the
positive frequency radiation mode function $\psi^{rad}_L(t)$ to the
linear combination $ \alpha^{mat}_L \psi^{mat}_L(t) + \beta^{mat}_L
\psi^{*mat}_L $ at time $t_{\rm eq}$.  One obtains \begin{eqnarray}
\label{MatterBog} \alpha^{mat}_L & =  &((2L+1)(2L+3))^{-1/2} \biggl(
-2 i(L+1) + \sqrt{W-1} + {iW \over 4 (L+1)} \biggr)\\ \beta^{mat}_L &
= & {-i \over 4} (L+1)^{-1}  ((2L+1)(2L+3))^{-1/2} W e^{-2 i (L+1)
t_{\rm eq}}. \nonumber \end{eqnarray} Here the constant $W$ is given
by \begin{equation} W= {\Omega_0 (1+Z_{\rm eq}) \over \Omega_0-1}.
\label{WDef} \end{equation} We are now in a position to evaluate the
multipole moments $ {\langle a_l^2 \rangle } $ of the angular
correlation function $C(\gamma)$.

\section{Multipole Moments of $C(\gamma)$} \label{section4} Combining
the results of the previous section, one can easily obtain a formula
for the multipole moments of the angular correlation function
$C(\gamma)$.  One replaces the metric perturbation that appears in
the Sachs-Wolfe formula (\ref{SachsWolfeFormula}) with expansion
(\ref{FieldOp}) of the field operator.  The resulting operator
depends upon an angle $\Omega$ on the celestial sphere.  One then
takes the expectation value of this operator with an identical
operator at a different point $\Omega'$ on the celestial sphere.
This yields the correlation function \begin{eqnarray} &C(\gamma)
\equiv \bigg\langle 0\bigg| {{\delta T} \over T} (\Omega){{\delta
T}\over T}(\Omega')\bigg|0\bigg\rangle=& \\ &{1 \over 4} \int_{t_{\rm
ls}}^0 dt \int_{t_{\rm ls}}^0 dt' \sum_{Llm} \sum_{L'l'm'} \dot
\psi_L(t) \dot \psi^*_L(t') {R_L^l}(|t|) {R_{L'}^{l'}}(|t'|) Y_{lm}
(\Omega) Y^*_{l'm'} (\Omega') \bigg\langle 0\bigg| c_{Llm}
c^\dagger_{L'l'm'}\bigg|0\bigg\rangle. &\nonumber \end{eqnarray} Here
$\gamma$ is the angle between the points $\Omega$ and $\Omega'$ on
the celestial sphere. Because the Sachs-Wolfe formula
(\ref{SachsWolfeFormula}) involves the time-derivative of the mode
function, we have defined $\dot \psi_L(t) \equiv d\psi_L(t)/dt$,
where $\psi_L$ is the mode function during the matter-dominated
epoch, given in (\ref{FinalModeFunction}).

To simplify this expression, first note that the matrix element
$\bigg\langle 0\bigg| c_{Llm} c^\dagger_{L'l'm'}\bigg|0\bigg\rangle
= \delta_{LL'}\delta_{ll'}\delta_{mm'}$.  This eliminates the triple
sum $\sum_{L'l'm'}$.  Because the summand is independent of the
summation index $m$, one may then explicitly carry out the sum over
$m$ using the addition formula for spherical harmonics, equation
(3.62) of reference \cite{Jackson}:  \begin{equation} \sum_{m=-l}^l
Y_{lm}(\Omega) Y^*_{l'm'} (\Omega') = {2l+1 \over 4 \pi} P_l(\cos
\gamma).  \end{equation} Because the argument of the Legendre
function $P_l(z)$ is the cosine of the angle $\gamma$ between the
points on the celestial sphere, this shows explicitly that the
correlation function depends only upon $\gamma$.  \begin{equation}
C(\gamma) = {1 \over 4} \sum_{L=2}^\infty \sum_{l=2}^L {2l+1 \over 4
\pi} P_l(\cos \gamma) \int_{t_{\rm ls}}^0 dt \int_{t_{\rm ls}}^0 dt'
\dot \psi_L(t) \dot \psi^*_L(t') {R_L^l}(|t|) {R_L^l}(|t'|)
\end{equation} Comparing this to the definition of the multipole
moments (\ref{cofgamma}), and noting that the summation
$\sum_{L=2}^\infty \sum_{l=2}^L $ is equivalent to the summation $
\sum_{l=2}^\infty \sum_{L=l}^\infty$, one immediately obtains a
simple formula for the multipole moment, \begin{equation} {\langle
a_l^2 \rangle } = {1 \over 4} \sum_{L=l}^\infty \int_{t_{\rm ls}}^0
dt \int_{t_{\rm ls}}^0 dt' \dot \psi_L(t) \dot  \psi^*_L(t')
{R_L^l}(|t|) {R_L^l}(|t'|) = {1 \over 4} \sum_{L=l}^\infty |\alpha_L
{I_L^l} + \beta_L {I_L^l}^* |^2.  \end{equation} The complex quantity
${I_L^l}$ is what remains of the integral of the mode function along
the radial null geodesic path.  \begin{equation} {I_L^l} \equiv
\int_{t_{\rm ls}}^0  dt {R_L^l}(|t|) {d \over dt} \psi^{mat}_L(t)
\end{equation} Note that we have assumed (as is implied in Table
\ref{FreeParameterTable}) that the surface of last scattering lies
within the matter-dominated epoch; the positive frequency mode
function during the matter phase is given by (\ref{PosFreqMat}).  The
Bolgoliubov coefficients are given by (\ref{alphabetadefinition})
\begin{equation} \alpha_L = \alpha_L^{rad} \alpha_L^{mat} +
\beta_L^{rad} \beta_L^{*mat} \quad {\rm and} \quad \beta_L =
\alpha_L^{rad} \beta_L^{mat} + \beta_L^{rad} \alpha_L^{*mat},
\label{BogDef} \end{equation} where the Bolgoliubov coefficients for
the matter and radiation transitions are defined by (\ref{MatterBog})
and (\ref{RadiationBog}).  In the next section, we discuss how the
multipole moments ${\langle a_l^2 \rangle }$ may be rapidly evaluated
using numerical techniques.


\section{Numerical Evaluation Of The Multipole Moments}
\label{NumEval} As discussed at the end of Section \ref{section2}, it
is convenient to define dimensionless quantities $ M_l \equiv
{\rho_{\rm Planck} \over \rho_{\rm de Sitter}} { l(l+1) \over 6}
   {\langle a_l^2 \rangle }$.  Using the previous formulae one may
   write this in the dimensionless form \begin{equation} M_l = {32
\pi^2 \over 3} {l(l+1) \over 6} {1+Z_{\rm eq} \over (1+Z_{\rm
end})^4} \biggl( {\Omega_0-1 \over \Omega_0} \biggr)^3
\sum_{L=l}^\infty  {1 \over L(L+1)(L+2)} |\alpha_L {J_L^l} + \beta_L
{J_L^l}^* |^2.  \end{equation} where \begin{eqnarray} &{J_L^l}
\equiv  \int_{t_{\rm ls}}^0  dt {R_L^l}(|t|) \csc^2(t/2 + B)
e^{-i(L+1)t} \times & \\ &\biggl[ -3 i (L+1) \cot(t/2 + B) - {3 \over
2} \csc^2(t/2 + B) + 2L^2 + 4L +1) \biggr] .& \nonumber
\end{eqnarray} Taken together with the definitions of $\alpha_L$ and
$\beta_L$ given (\ref{BogDef}), (\ref{MatterBog}), and
(\ref{RadiationBog}), the constants $Q$ and $W$ defined in
(\ref{QDef}) and (\ref{WDef}), and the radial function
${R_L^l}(\chi)$ defined in (\ref{RadialFun}) this is a self-contained
formula for calculating $M_l$.

Before discussing the evaluation of $M_l$ in general, it is worth
commenting on two limits.  The first limit is the $Z_{\rm end} \to
\infty$ case, where the amount of inflation is large.  In this case,
it is easy to see that $Q$ and hence $\alpha_L$ and $\beta_L$ diverge
$\propto (1+Z_{\rm end})^2$.  Thus in the limit, $M_l$ converges.  A
second interesting limit is the spatially-flat one, $\Omega_0-1 \to
0^+$, where the density parameter approaches unity from above.  In
this case, it is easy to see that $Q$ and hence $\alpha_L$ and
$\beta_L$ diverge as $(\Omega_0-1)^{-1}$, and the integral ${J_L^l}$
diverges as $(\Omega_0-1)^{-1/2}$.  Once again, the limit is
well-defined.  In addition, in this case, the sum over $L$ can be
re-written as an integral, recovering the $k=0$ spatially-flat
formula given in \cite{AllenKoranda}.

We evaluated $M_l$ using an fourth-order Runge-Kutta adaptive
stepsize integrator \cite{PressEtAl} to obtain the integral which
defines $J_L$.  In cases of interest, one frequently needs to include
many values of $L$ in the summation.  In practice we found that
summing over the range $L \in l,l+1,\cdots,l_{\rm max}$ with $l_{\rm
max}=32 + (5l+10)/|t_{\rm ls}|$ gave results accurate to a few
percent for reasonable ranges of the free parameters listed in Table
\ref{FreeParameterTable}.  Rather than compute the ${J_L^l}$ one at a
time, it is more practical to compute them ``en masse'', determining
$J_l^l, J_{l+1}^l, \cdots, J_{l_{\rm max}}^l$ simultaneously.  This
can be done easily because the associated Legendre functions may be
computed with a stable upwards recursion relation.

\subsection{Second-Order Recursion Relations} The upwards recursion
relation for the associated Legendre functions is given in equation
(8.731.2) of reference \cite{GradshteynRyzhik}.  \begin{equation}
{\rm P}^{-(l+1/2)}_{j+l+1/2}(z) = {2 (l+j) \over 2 l + j + 1} z {\rm
P}^{-(l+1/2)}_{(j-1)+l+1/2}(z) + {1-j \over 2 l + j + 1} {\rm
P}^{-(l+1/2)}_{(j-2)+l+1/2}(z) \quad {\rm for} \quad j=2,3,\cdots,
\label{RecursionRelation} \end{equation} together with the boundary
conditions (or initial values) given in equation (8.755.1) of
reference \cite{GradshteynRyzhik}:  \begin{equation} {\rm
P}^{-(l+1/2)}_{l+1/2}(\cos \chi) = {1 \over \Gamma(l+3/2)} \biggl(
{\sin \chi \over 2} \biggr)^{l+1/2} \quad {\rm and} \quad {\rm
P}^{-(l+1/2)}_{l+3/2}(z) = z {\rm P}^{-(l+1/2)}_{l+1/2}(z).
\end{equation} These relations may be used to obtain a recursion
relation and initial values for the radial functions ${R_L^l}$.  The
initial values are \begin{equation} {R_l^l}(\chi) = \sqrt{  (l-1)
\Gamma(l+1) \over 2 \sqrt{\pi} \Gamma(l+3/2)} (\sin \chi)^{l/2-1}
\quad {\rm and} \quad {R_{l+1}^l}(\chi) = \cos \chi \sqrt{ 2 l(l+1)
\over l+3} {R_l^l}(\chi) \end{equation} and the recursion relation is
obtained from (\ref{RecursionRelation}) \begin{eqnarray} &{\rm for}
\quad j=2,3,\cdots \quad {R_{l+j}^l}(\chi) = &\\ &
\sqrt{4(l+j)^2(l+j-1) \over j (l+j+2)(2l+j+1)} \cos \chi
{R_{l+j-1}^l}(\chi) - \sqrt{(j-1) (2 l+j)(l+j-1)(l+j-2) \over
j(2l+j+1)(l+j+1)(l+j+2)} {R_{l+j-2}^l}(\chi) & \nonumber
\end{eqnarray} Although this recursion relation does not appear to be
stable, our experience has been that it accurately determines
${R_L^l}$ for $l \le L \le l+6000$.

\section{Numerical Results and Conclusions} \label{section6} The
numerical results are presented as a series of graphs Figs.
\ref{figure1}, \ref{figure2}, and \ref{figure3} showing the values of
$ M_l \equiv {\rho_{\rm Planck} \over \rho_{\rm de Sitter}} { l(l+1)
\over 6} {\langle a_l^2 \rangle }$.  For all of these graphs, we have
taken $Z_{\rm end}=10^{26}$, $Z_{\rm eq}=10^4$ and $Z_{\rm ls}=1300$,
and varied the density parameter $\Omega_0$.  The graphs also show
the values of $M_l$ for the spatially-flat $k=0$ case, taken from
\cite{AllenKoranda}.  This case corresponds to the critically-bound
$\Omega_0 \to 1$ limit.

It is clear from the figures that this limit is quickly approached;
when $\Omega_0=1.1$ the $M_l$ are almost indistinguishable from the
$k=0$ spatially-flat case.  It is not hard to see why.  The effects
of the spatial curvature only appear if the past light cone of the
observer, taken back to the surface of last scattering, actually
``probes'' a substantial fraction of the spatial three-sphere.  If
the past light cone fails to do this, then within the past light cone
the universe is indistinguishable (to good approximation) from a
spatially-flat model.

The fraction of the three-sphere ($S^3$) within this past light cone
is easy to determine.   The three-volume contained within angle
$\chi_{\rm max}$ from the point $\chi=0$ of the unit-radius $S^3$ may
be obtained by integrating $\sqrt{\det P_{ij}}$ where $P_{ij}$ is the
three-metric (\ref{UnitSphereMetric}).  One obtains $V(\chi_{\rm
max})=  \pi (2 \chi_{\rm max} - \sin 2 \chi_{\rm max})$.  The total
volume of $S^3$ is $V(\pi)=2 \pi^2$.  If we assume that $Z_{\rm ls}$
is much larger than one, then the fraction of the volume of $S^3$
contained within the past light cone is approximately
\begin{equation} f(\Omega_0) \equiv { V(|t_{\rm ls}|) \over V(\pi)}
\approx { V(2 \arcsin \sqrt{\Omega_0-1 \over \Omega_0} ) \over 2
\pi^2}.  \end{equation} For $\Omega_0$ near 1, this fraction is
well-appoximated by \begin{equation} f(\Omega_0) \approx {16 \over 3
\pi} (\Omega_0-1)^3.  \end{equation} Thus when $\Omega_0=1.1$ the
past light cone only explores about $1/1000$ of the spatial volume.
Even if $\Omega_0=2$ the fraction of the three-sphere that is
observed is only $f(\Omega_0=2)={1 \over 3}-{ \sqrt{3} \over 4 \pi}
\approx 0.1955\cdots$.  This is why the multipole moments are not
very sensitive to $\Omega_0$ provided it is close to unity.

The extension of this calculation to the case of a spatially open FRW
universe appears straightforward.  However it turns out to be much
more difficult than expected, primarily because the correct choice of
initial state is not the obvious one, and because the final result
for the multipole moments appears to contain logarithmic (infra-red)
divergences at zero frequency.  The spatially open case will be the
subject of a forthcoming paper.

\acknowledgments This research work has been partially supported  by
National Science Foundation grant number PHY91-05935.  The work of RC
is supported by the NASA through Grant No. NAGW-2381 (at Fermilab).



\begin{figure} \caption{The normalized multipole moments $ M_l \equiv
{\rho_{\rm Planck} \over \rho_{\rm de Sitter}} { l(l+1) \over 6}
   {\langle a_l^2 \rangle } $ of the CBR temperature fluctuations are
   shown as a function of the multipole number $l$, for a
spatially-flat ($\Omega_0=1$) and for spatially-closed ($\Omega_0
>1$) cosmological models.  $\Omega_0-1$ needs to be fairly large for
the effects of the spatial curvature to be significant.  The models
being compared all have cosmological parameters defined by the
redshifts $Z_{\rm ls}=1300$, $Z_{\rm eq}=10^4$ and $Z_{\rm
end}=10^{26}$.  } \label{figure1} \end{figure}

\begin{figure} \caption{The normalized multipole moments $ M_l \equiv
{\rho_{\rm Planck} \over \rho_{\rm de Sitter}} { l(l+1) \over 6}
{\langle a_l^2 \rangle } $ are shown as a function of $\Omega_0-1$
for $l=2,5,10,20,30,50$.  In all cases, the models being compared
have the same cosmological parameters as in
Fig.\ \protect\ref{figure1}.  Only when $\Omega_0$ becomes
significantly larger than one do the multipole moments change
significantly from the spatially-flat case.  } \label{figure2}
\end{figure}

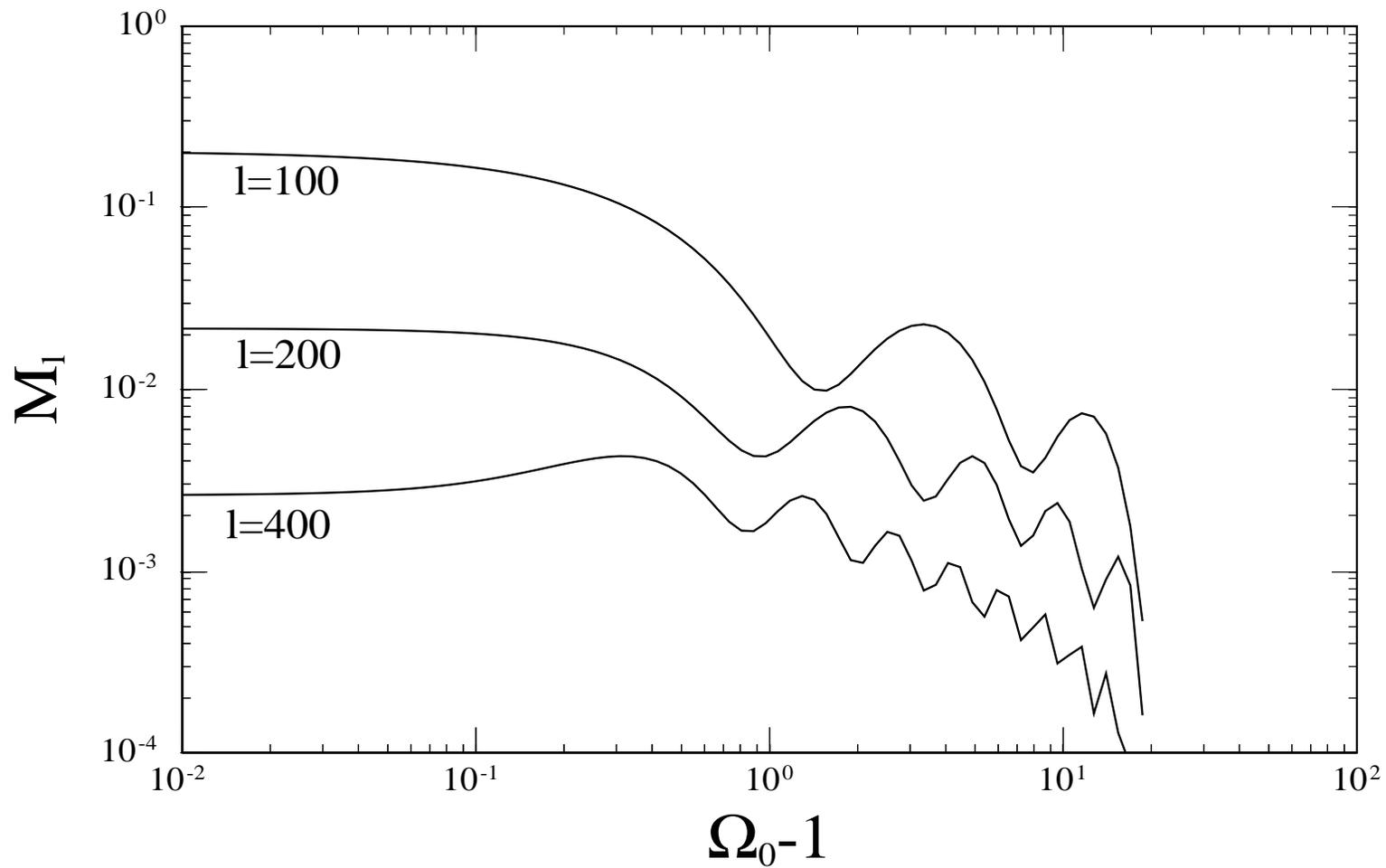
\begin{figure} \caption{The normalized multipole moments $ M_l \equiv
{\rho_{\rm Planck} \over \rho_{\rm de Sitter}} { l(l+1) \over 6}
{\langle a_l^2 \rangle } $ are shown as a function of $\Omega_0-1$
for $l=100,200,400$.  In all cases, the models being compared have
the same cosmological parameters as in Figs.\ \protect\ref{figure1}
and \protect\ref{figure2}.  Only when $\Omega_0$ becomes
significantly larger than one do the multipole moments change
significantly from the spatially-flat case.} \label{figure3}
\end{figure}


\begin{table} \caption{List of the free parameters that define the
cosmological model.} \begin{tabular}{ccccc}
Parameter&Units&Range&Description\\ \tableline $H_0$&$\rm
length^{-1}$&$H_0>0$&Present-day Hubble expansion rate\\
$\Omega_0$&dimensionless&$\Omega_0>1$&Present-day density parameter\\
$Z_{\rm ls}$&dimensionless&$Z_{\rm ls}>0$&Redshift at last scattering
of CBR\\ $Z_{\rm eq}$&dimensionless&$Z_{\rm eq}>Z_{\rm ls}$&Redshift
at equal matter/radiation energy density\\ $Z_{\rm
end}$&dimensionless&$Z_{\rm end}>Z_{\rm eq}$&Redshift at end of de
Sitter inflation\\ \end{tabular} \label{FreeParameterTable}
\end{table}
\end{document}